\documentclass[aps,pre,onecolumn,preprintnumbers,amsmath,amssymb,10pt,a4paper,nofootinbib]{revtex4}
\usepackage{geometry}
\pdfoutput=1
\usepackage[dvips]{graphicx}
\usepackage{amssymb,amsfonts,amsmath}
\usepackage{color}
\usepackage{eurosym}
\usepackage{latexsym}
\usepackage{amssymb}
\usepackage{amsfonts}
\usepackage{amsmath}
\usepackage{theorem}
\usepackage{graphicx}
\usepackage{setspace}
\usepackage{graphicx}% Include figure files
\usepackage{dcolumn}% Align table columns on decimal point
\usepackage{bm}% bold math
\usepackage{subfigure}% allow subfigures
\usepackage{amsmath}
\usepackage{amsfonts}
\usepackage{multirow}
\newcommand{\be}{\begin{equation}}
\newcommand{\ee}{\end{equation}}
\newcommand{\bd}{\begin{displaymath}}
\newcommand{\ed}{\end{displaymath}}
\newcommand{\BE}{\begin{eqnarray}}
\newcommand{\EE}{\end{eqnarray}}

\newcommand{\bx}{\ensuremath{\mathbf{x}}}
\newcommand{\by}{\ensuremath{\mathbf{y}}}
\newcommand{\bz}{\ensuremath{\mathbf{z}}}

\newcommand{\olx}{\overline{x}}
\newcommand{\oly}{\overline{y}}

\newcommand{\boldxi}{{\mbox{\boldmath $\xi$}}}
\newcommand{\boldeta}{{\mbox{\boldmath $\eta$}}}

\newcommand{\boldphi}{{\mbox{\boldmath $\varphi$}}}

\newcommand{\bzeta}{{\mbox{\boldmath $\zeta$}}}

\newcommand{\avg}[1]{\left\langle{#1}\right\rangle}

\begin{document}

 \title{ Cycles of cooperation and defection in imperfect learning}
% \thanks[label1]{}
\author{Tobias Galla}
\email{tobias.galla@manchester.ac.uk}
% \ead[url]{home page}
% \thanks[label2]{}
% \corauth[cor1]{}
\address{Theoretical Physics Group, School of Physics and Astronomy, \\
The University of Manchester, Manchester M13 9PL, United Kingdom}

%\sffamily

%\begin{article}

\begin{abstract}
When people play a repeated game they usually try to anticipate their opponents' moves based on past observations, and then decide what action to take next. Behavioural economics studies the mechanisms by which strategic decisions are taken in these adaptive learning processes. We here investigate a model of learning the iterated prisoner's dilemma game. Players have the choice between three strategies, always defect (ALLD), always cooperate (ALLC) and tit-for-tat (TFT). The only strict Nash equilibrium in this situation is ALLD.  When players learn to play this game convergence to the equilibrium is not guaranteed, for example we find cooperative behaviour if players discount observations in the distant past. When agents use small samples of observed moves to estimate their opponent's strategy the learning process is stochastic, and sustained oscillations between cooperation and defection can emerge. These cycles are similar to those found in stochastic evolutionary processes, but the origin of the noise sustaining the oscillations is different and lies in the imperfect sampling of the opponent's strategy. Based on a systematic expansion technique, we are able to predict the properties of these learning cycles, providing an analytical tool with which the outcome of more general stochastic adaptation processes can be characterised. \end{abstract}

\maketitle
%\keywords{Evolutionary Dynamics | multiplayer games | multiple strategies}

\section{Introduction}

The mathematical theory of games goes back to von Neumann and Morgenstern \cite{Neumann1953}, and was initially concerned with the study of equilibrium points \cite{Nash1950,Nash1951}. The idea that players would be able to compute such equilibria requires severe assumptions, in particular perfect rationality and full knowledge of the game. Additionally each player has to assert that all other players are rational as well. Von Neumann and Morgenstern stress the limitations of their approach explicitly: {\em `We repeat most emphatically that our theory is thoroughly static. A dynamic theory would unquestionably be more complete and therefore preferable.'} \cite{Neumann1953}. 

Since the work of von Neumann and Morgenstern more than 70 years ago, several different routes have been taken to formulate a dynamical theory of games. Evolutionary game theory was launched by Maynard Smith in the 1970s and considers time-dependent dynamics of populations of players \cite{MaynardSmith1973,MaynardSmith1998}. Each individual in the population carries a pure strategy, inherited from its parent(s), and agents then reproduce and pass on their strategies to their offspring, with a reproduction rate depending on the performance in the game. The strategic content of the population evolves, with the concentration of successful strategies increasing over time, and those of less successful strategies being reduced. Evolutionary game theory has been used to model a vast number of phenomena in the social sciences and in economics \cite{Nowak2001, Myerson1991,Hens05,hofbauer,Gintis2000,Vega2003,nowak}. 

These applications include in particular the study of the emergence of cooperation and altruism \cite{Sigmund2010}. The evolution of cooperative behaviour under selection pressure constitutes a formidable puzzle. The dynamics of evolution is governed by a fierce competition between individuals, and only those who act in their own interest and who selfishly promote their own evolutionary success at the expense of their competitors should prevail in the long-run. Nevertheless altruism and cooperative behaviour are found in a number of evolved systems, ranging from cooperating genes or cells to cooperating animals or humans in social contexts \cite{axelrod,axelrod2,nowak5}. The question how cooperative behaviour has evolved under strictly competitive and selective dynamics is still unresolved, and has recently been listed as one of the 125 big open problems in science \cite{pennisi}.

Our goal here is to address the emergence of cooperation in a third approach to game theory. We focus on adaptive learning processes of a small fixed set of individuals, who interact repeatedly in a game \cite{fudenberg,CesaBianchi2006,Young2004,camerer1,Camerer2003,Ho2007}. Players observe their opponents' actions and aim to react dynamically by adapting their own strategic propensities, learning from past experience. Such learning models are of particular importance for the understanding of experiments in behavioural game theory, where human subjects play a given game repeatedly under controlled conditions, see e.g. \cite{camerer1,Camerer2003,Henrich2004, Ho2007, Semmann2003,Traulsen2010}.
A-priori it is not clear whether adaptation will converge to Nash equilibria. Learning has for example been seen to fail to converge in games with cyclic payoff structures, and complex trajectories including limit cycles, quasiperiodic motion and Hamiltonian chaos have instead been identified \cite{satopre, satopnas,satophysica}. 
 
Mathematical models of cooperative behaviour are often based on stylised games played by a small number of interacting individuals, each choosing from a small number of strategies. Such games have been characterised as {\em `mathematical x-ray[s] of crucial features'} of real-word situations \cite{Camerer2003}. The most basic setup is the celebrated prisoners' dilemma, a game in which two players have the choice between cooperation and defection.  Defection dominates cooperation in this game, no matter what the other player decides to do, either player will always do better defecting than cooperating. Fully rational players hence end up playing the only equilibrium strategy, defection, and have to put up with the a suboptimal payoff, when they could have scored higher had they both cooperated.

If the prisoners' dilemma is iterated, more complex behaviour is possible and the space of all strategies grows rapidly as the number of iterations is increased. In order to make progress it is therefore necessary to restrict  the mathematical analysis to a subset of this space. We will focus on three strategies: always defect (ALLD), always cooperate (ALLC) and tit-for-tat (TFT). Players using the TFT strategy
cooperate in the first iteration and then proceed by playing whatever the opponent played in the previous round. The replicator-mutator dynamics of populations of players engaging in this game have been studied in \cite{imhof,bladon}.  ALLD has been identified as the deterministic replicator fixed point, and mutation has been seen to move the attractor toward cooperation. Demographic noise in finite populations can alter the dynamics and can induce coherent evolutionary cycles between defection and cooperation.

As one main result we show that the effects of memory-loss in the learning dynamics are very similar to those of mutation in evolutionary dynamics. While deterministic learning in the absence of memory loss converges to ALLD, this Nash equilibrium is no longer an attractor when players discount observations in the distant past, and a different fixed point, involving all three pure strategies, emerges. Deterministic replicator-type equations are a faithful description of the learning process if and only if a large number of observations of the opponent's actions is made before players update their own strategic preferences. If, on the contrary, adaptation occurs more frequently and is based only a small sample of observations, the dynamics becomes stochastic. The source of randomness lies in the imperfect sampling of the opponent's mixed strategy profile. When each player uses a small number of observed actions to estimate the opponent's mixed strategy, then the estimate will generally be subject to statistical errors. The observed actions were chosen according to the opponent's mixed strategy profile, but still they are random variables. This source of noise different from the origin of demographic noise in the evolution of finite populations. Nevertheless the effects are similar: as our second main result we show that sustained cycles between cooperation and defection can emerge in stochastic learning, similar to those found in evolutionary scenarios of the iterated prisoner's dilemma game \cite{imhof}. We are able to predict the characteristic frequency and power spectra of these cycles analytically as a function of the parameters of the game and the learning dynamics.

\section{Model}
To define the iterated prisoner's dilemma we will follow the notation of \cite{imhof}. Assuming that $m$ iterations of the prisoner's dilemma are played in any one interaction of the two players, and that a complexity cost $c$ is associated with playing TFT the payoff matrix is given by 
\vspace{2em}
 \begin{center}
\begin{tabular}{|l|c|c|c|}
\hline
& ALLC& ALLD&TFT\\\hline
ALLC & $R$ & $S$ & $R$ \\
ALLD & $T$ & $P$ & $\frac{T+P(m-1)}{m}$ \\
TFT & ~~$\frac{Rm-c}{m}$~~ &~~ $\frac{S+P(m-1)-c}{m}$ ~~&~~ $\frac{Rm-c}{m}$ ~~ \\
\hline
\end{tabular}~~~,
\end{center}
\vspace{2em}
i.e. a player playing ALLC will for example receive a payoff of $R$ (per round) when meeting another ALLC player, a payoff of $S$ when playing against ALLD and a payoff of $R$ upon encountering TFT.  We will denote the payoff matrix elements as $a_{ij}$, where $i,j=1,2,3$ label the strategies ALLC, ALLD and TFT respectively. Throughout this paper we use $T=5, R=3, P=1, S=0.1, m=10, c=0.8$.

In our model the game is played repeatedly by two players Alice and Bob. We will assume that Alice carries a (time-dependent) mixed strategy profile $\bx(t)=(x_1(t),x_2(t),x_3(t))$ and similarly Bob's mixed strategy profile at $t$ is  $\by(t)=(y_1(t),y_2(t),y_3(t))$. We will write $i(t)$ for Alice's action at time $t$, and $j(t)$ for Bob's action, i.e. $i(t),j(t)\in\{\mbox{ALLC, ALLD, TFT}\}$. Following \cite{Camerer2003,camerer1, Ho2007} each player keeps attractions for each of the pure strategies. Alice's attractions at time $t$ are labelled by $A_i(t)$ and Bob's attractions by $B_j(t)$. We will again follow \cite{Camerer2003,camerer1,Ho2007} as well as \cite{satopnas,satopre,satophysica} and assume that attractions determine choice probabilities through a logit rule, i.e. that the probabilities for Alice and Bob to play the different pure strategies at time $t$ are given by
\be
x_i(t)=\frac{e^{\beta A_i(t)}}{\sum_{k}e^{\beta A_k(t)}}, ~~~~~  y_j(t)=\frac{e^{\beta B_j(t)}}{\sum_{k}e^{\beta
B_k(t)}}\label{eq:probxy}.
\ee 
The variable $\beta$ is a model parameter, and describes the intensity of selection or response sensitivity \cite{Ho2007}. For $\beta\to \infty$ the players strictly choose the pure action with highest attraction, for $\beta=0$ they play at random. We will here restrict the discussion to models in which both players use the same intensity of selection, generalisation to heterogeneous intensities is straightforward.

A simple re-inforcement learning dynamics is then defined by the following update rules for the attractions
\BE
A_k(t+1)&=&(1-\lambda)A_k(t)+a_{k,j(t)}, \nonumber \\
B_k(t+1)&=&(1-\lambda)B_k(t)+a_{k,i(t)}.\label{eq:abupdate}
\EE
Alice's attraction $A_k$ is therefore re-inforced by the payoff $a_{k,j(t)}$ she would have received at time $t$ had she played action $k$, and similarly for Bob. The parameter $\lambda$ indicates memory loss, observations in the distant past carry a lesser weight than more recent rounds. For $\lambda=0$ the players have perfect memory of past play, and use the outcome of all past rounds with equal weight to determine their attractions. In particular $A_k$ for example is then the total payoff Alice would have received had she always played action $k\in\{\mbox{ALLC, ALLD, TFT}\}$, given Bob's moves. For $\lambda>0$ experiences in the past are discounted exponentially. This may happen voluntarily as part of a learning mechanism or simply be due to fading memories and limited mental capacities. We will occasionally refer to $\lambda$ as a memory-loss rate or discounting factor. We assume that both players learn at identical memory-loss rates, generalisation to heterogeneous learning rules ($\lambda_{\mbox{\footnotesize Alice}}\neq\lambda_{\mbox{\footnotesize Bob}}$) is straightforward. Up to relabelling this learning rule is a special case of experience-weighed attraction learning, as discussed in \cite{Camerer2003, Ho2007}. More general learning dynamics are discussed in the appendix.

The process defined by Eqs. (\ref{eq:probxy},\ref{eq:abupdate}) is intrinsically stochastic, the actions $i(t)$ and $j(t)$ are drawn from the mixed strategy profiles $\bx(t)$ and $\by(t)$ respectively, and accordingly the attractions $A_k(t)$ and $B_k(t)$ are random variables as well.  Simple averaging, taking into account that $i(t)$ takes the value $i(t)=\ell$ with probability $x_\ell(t)$ and that $j(t)=\ell$ with probability $y_\ell(t)$, results in the following average attraction update
\BE
A_k(t+1)&=&(1-\lambda)A_k(t)+\sum_{\ell=1}^3~a_{k\ell}y_\ell(t), \nonumber \\
B_k(t+1)&=&(1-\lambda)B_k(t)+\sum_{\ell=1}^3~a_{k\ell}x_\ell(t).\label{eq:meanabupdate}
\EE
Limiting dynamics of this type can provide insight into the expected outcome of learning. Deterministic learning has been shown to lead to modified replicator equations in a continuous-time limit \cite{satopre,satopnas,satophysica}. Analyses of discrete-time deterministic learning can be found in \cite{ochea}. The derivation the deterministic dynamics relies on an adiabatic approximation though, it is assumed that strategy updates occur on a much slower time scale than the actual play. In order to perform the update of Eq. (\ref{eq:meanabupdate}) Alice has to have full knowledge of Bob's mixed strategy $\by(t)$, and Bob needs to be aware of Alice's strategy $\bx(t)$. This will generally be very hard to achieve for the players. Eqs. (\ref{eq:meanabupdate}) are therefore only an approximate description of the learning process, and can at best be expected to describe the average behaviour.  Describing learning in terms of these deterministic equations is procedurally akin to describing the average behaviour of evolving populations by means of deterministic replicator equations.  To understand the nature of the approximation underlying the deterministic limit it is instructive to interpolate between the deterministic average process and the actual stochastic dynamics. We here consider a batch learning process, in which each player samples $N$ actions of their respective opponent, and then updates their attractions. The above `adiabatic' approximation consists in assuming stationarity of the mixed strategy profiles between attraction updates. Specifically we introduce the following process
\BE
A_k(\tau+1)&=&(1-\lambda)A_k(\tau)+\frac{1}{N}\sum_{\alpha=1}^N~a_{k,i_\alpha(\tau)}, \nonumber \\
B_k(\tau+1)&=&(1-\lambda)B_k(\tau)+\frac{1}{N}\sum_{\alpha=1}^N~a_{k,j_\alpha(\tau)}.\label{eq:abbatchupdate}
\EE
The interpretation of these update rules is as follows: at time $\tau$ Alice independently selects $N$ actions $i_\alpha(\tau)$ ($\alpha=1,\dots,N$) following her mixed strategy profile $\bx(\tau)$ at that time. I.e. the $\{i_\alpha(\tau)\}$ are independent random variables, and for each $\alpha$ one has $i_\alpha(\tau)=\ell$ with probability $x_\ell(\tau)$. Bob draws his actions $j_\alpha(\tau)$ in a similar manner, using his mixed strategy $\by(\tau)$.These actions represent the moves made by the two players in $N$ successive rounds of the game, the mixed strategies $\bx(\tau)$ and $\by(\tau)$ are kept fixed during the course of these rounds. At the end of the batch of $N$ rounds both Alice and Bob update their attractions based on Eq. (\ref{eq:abbatchupdate}), and then adapt their mixed strategy profiles using Eq. (\ref{eq:probxy}) (with $t$ replaced by $\tau$). We have intentionally used the notation $\tau$ rather than $t$ to denote time steps of this batch dynamics. One unit of time $\tau$ corresponds to $N$ repetitions of the game, i.e. to $N$ units of time $t$. We will refer to $N$, the number of observations made in between updates of the attractions, as the batch size, following the language of machine learning \cite{saad}. Small batch sizes $N$ correspond to fast adaptation. If $N=1$ we recover the original dynamics (\ref{eq:abupdate}) where strategy updates are performed
after every single round of the game. Large $N$ on the other hand indicate infrequent adaptation, the limit of infinite batches leads to the deterministic update rule Eq. (\ref{eq:meanabupdate}). This limit is based on the assumption that the mixed strategy profiles $\bx(\tau)$ and $\by(\tau)$ are stationary during each batch of $N$ repetitions of the game. This assumption will be irrelevant at small batch sizes $N$, but more severe in the limit of large $N$. Taking the limit $N\to\infty$ to derive the deterministic learning rule is analogous to the procedure leading to a description of evolving populations in terms of deterministic replicator equations. In evolutionary systems these descriptions are accurate for populations with an infinite number of individuals. Stochastic corrections cannot be neglected in finite populations, and the resulting noise has been seen to alter the dynamics substantially, see e.g. \cite{imhof,bladon}. Similarly, real-world players do not operate adiabatic learning dynamics, but instead small batch sizes $N$ are probably more appropriate to describe experiments in behavioural economics.  It is therefore important to go beyond the deterministic limit of Eq. (\ref{eq:meanabupdate}) and to study stochastic effects at finite batch sizes. First steps have been taken in \cite{galla}, and it is one of the main purposes of this work to apply these ideas to the iterated prisoner's dilemma game.

\section{Results and Discussion}
We illustrate the outcome of the continuous-time deterministic learning (see appendix) in Fig. \ref{fig:det}. At low memory-loss rates the dynamics is essentially governed by the standard replicator equations, and the system has a single stable fixed point near ALLD, similar to what is reported for low mutation rates in evolutionary systems \cite{imhof}. As the memory-loss rate is increased ALLD remains a stable attractor, but cyclic attractors around an unstable fixed-point emerge (top right panel of Fig. \ref{fig:det}). At even higher memory loss this second fixed point becomes a stable spiral. Provided players do not discount past play too strongly this spiral fixed point is located in the vicinity of the ALLC/TFT edge of the strategy simplex, and we conclude that moderate memory loss may enhance cooperative behaviour. When the memory becomes even shorter the fixed point moves towards the centre of the simplex. In the extreme case of full memory-loss $\lambda=1$ players ignore the past history beyond the last iteration entirely. Depending on the response sensitivity both players play essentially at random, the three strategies are used with very similar frequencies. 

It is interesting to note that the outcome of deterministic learning with memory-loss resembles the behaviour of replicator-mutator dynamics of this game \cite{imhof}.  Discounting past experience in learning and mutation in evolution both promote cooperation when they are moderate in strength. The attractors of learning with quick memory loss on the other hand are similar to those of evolutionary systems in which mutation dominates selection.

We will now move to learning at {\em finite} batch sizes $N$. Players are then no longer able to obtain a perfect sample of their opponents' mixed strategy profile before updating their own strategic propensities, and the dynamics becomes stochastic. Results of numerical simulations are shown in Fig. \ref{fig:timeseries}. We here focus on a regime in which deterministic learning approaches a fixed point. Stochastic learning at the same discounting rate and intensity of selection results in sustained cycles between cooperation and defection. The amplitude of these cycles is found to scale as $N^{-1/2}$ in the batch size, but the coefficient multiplying $N^{-1/2}$ can be substantial (see appendix) so that the oscillations can have a significant amplitude. The inset of Fig. \ref{fig:timeseries} confirms that average of several independent runs of the stochastic dynamics is accurately described by the deterministic update rules of Eqs. (\ref{eq:meanabupdate}).

The cycling behaviour of the stochastic learning process can be understood as the result of an amplification mechanism, which turns intrinsic white noise into coherent oscillations \cite{alan}. The intuitive picture is here as follows: at the memory-loss rate chosen in Fig. \ref{fig:timeseries} the deterministic dynamics spirals into a stable fixed point asymptotically, the relevant eigenvalue of the dynamics is complex. If an instantaneous perturbation were applied to the deterministic system at the fixed point, the dynamics would return to the fixed point following a trajectory of damped oscillations. At finite batch sizes, however, the dynamics is subject to persistent random fluctuations, constantly driving the system away from the fixed point. The combination of this permanent 'excitation' and the oscillatory relaxation results in a coherently maintained cyclic pattern. 

\begin{figure}[h]
\vspace{0em}
\begin{center}
\includegraphics[width=0.5\textwidth]{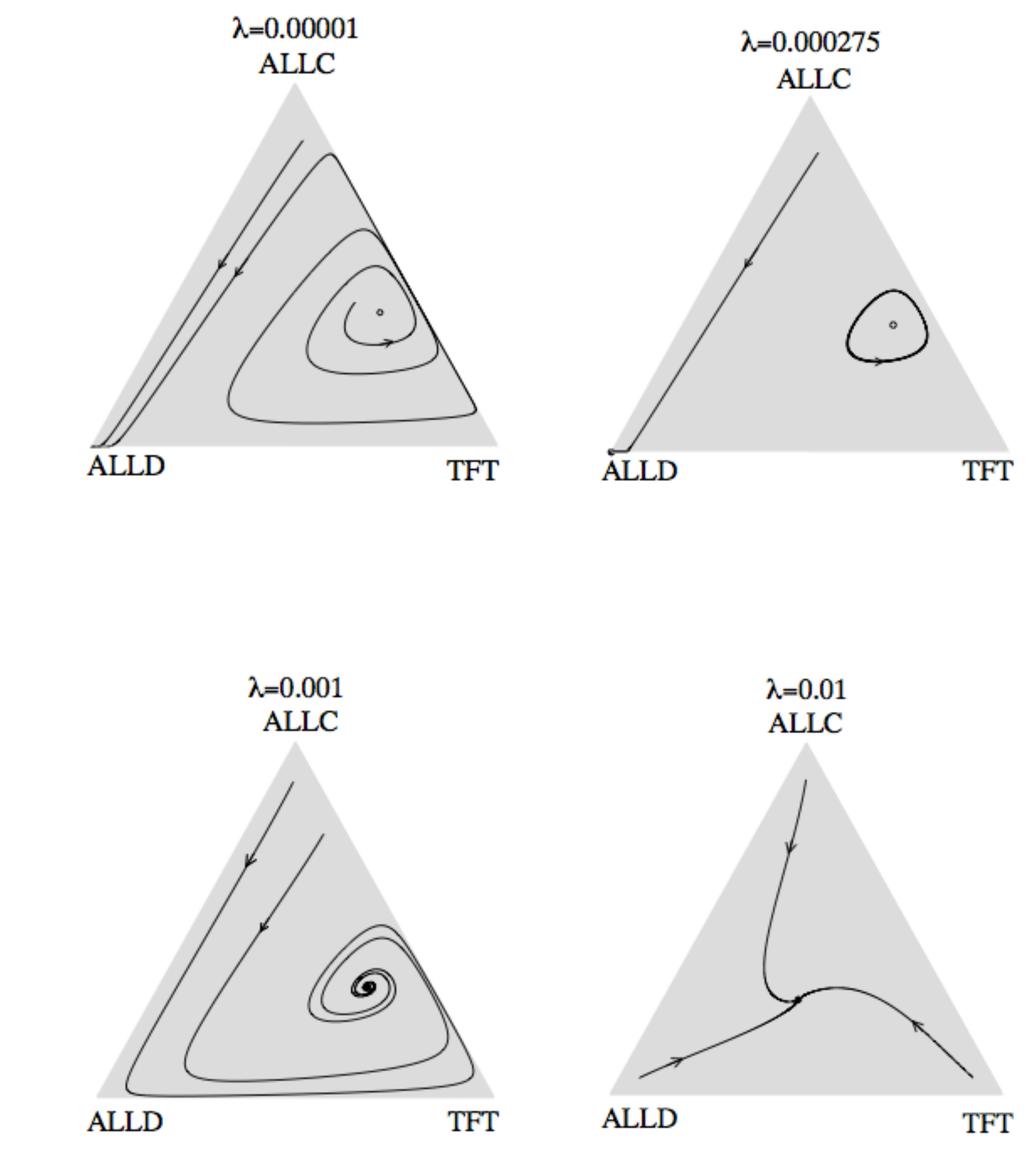}
\end{center}
% Here is how to import EPS art
\caption{\label{fig:det} Illustration of the behaviour of the deterministic continuous-time learning  (see appendix) at different memory-loss rates $\lambda$. Intensity of selection is $\beta=0.01$.}
\end{figure}

\begin{figure}[h]
\vspace{0em}
\begin{center}
\includegraphics[width=0.45\textwidth]{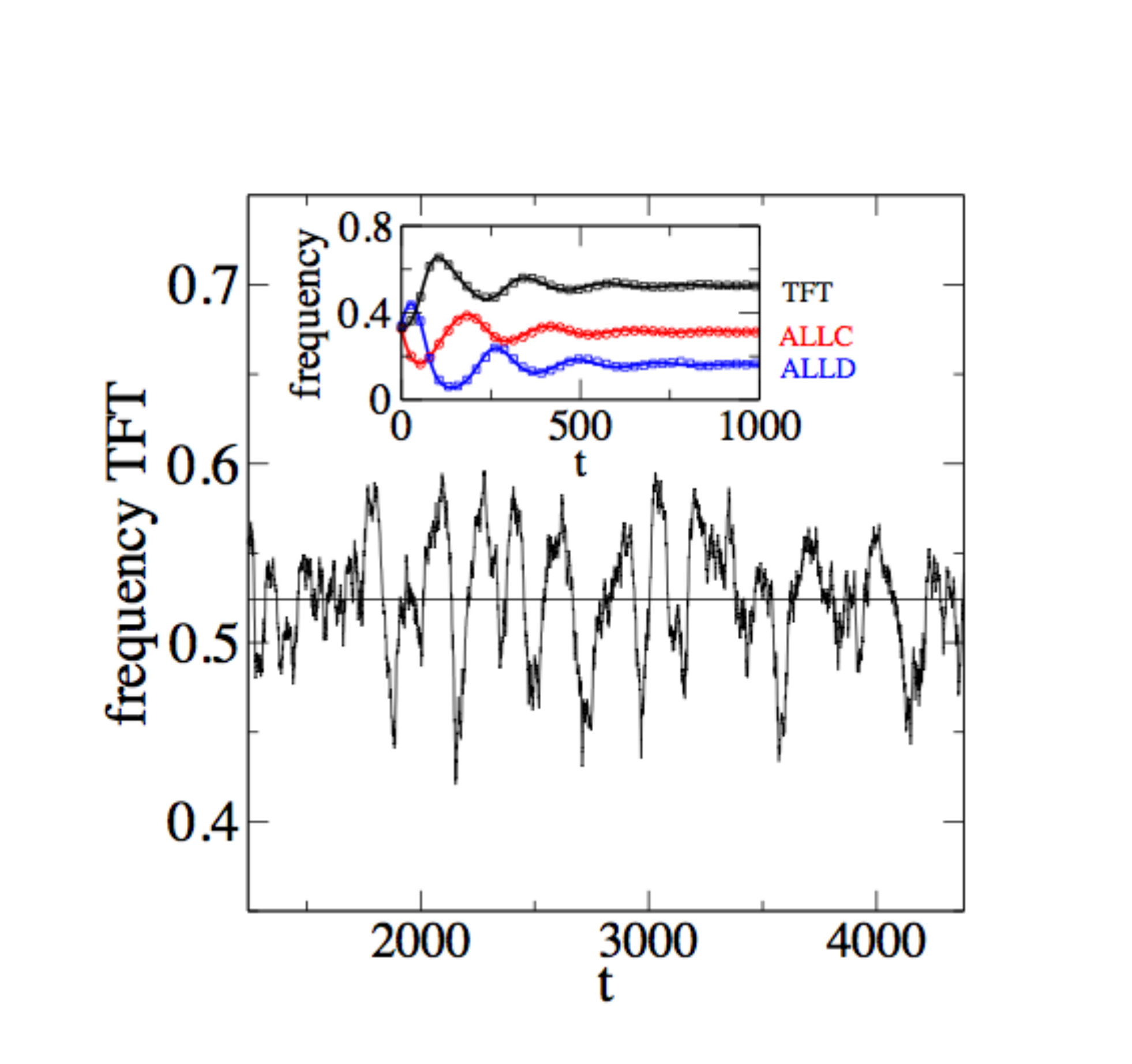}
\end{center}
% Here is how to import EPS art
\caption{\label{fig:timeseries} (Color on-line) Sustained oscillations in the stochastic dynamics. Frequency with which TFT is played by Alice as a function of time at $N=10$ observations between adaptation events. The horizontal line is the fixed point of deterministic learning. The inset shows the frequencies of ALLC, ALLD and TFT in the initial phase of the dynamics. Solid lines are the outcome of deterministic learning, symbols show data from an average over $100$ independent runs of stochastic learning at $N=10$. Model parameters are $\beta=0.1, \lambda=0.01$.}
\end{figure}

\begin{figure}[h]
\vspace{0em}
\begin{center}
\includegraphics[width=0.4\textwidth]{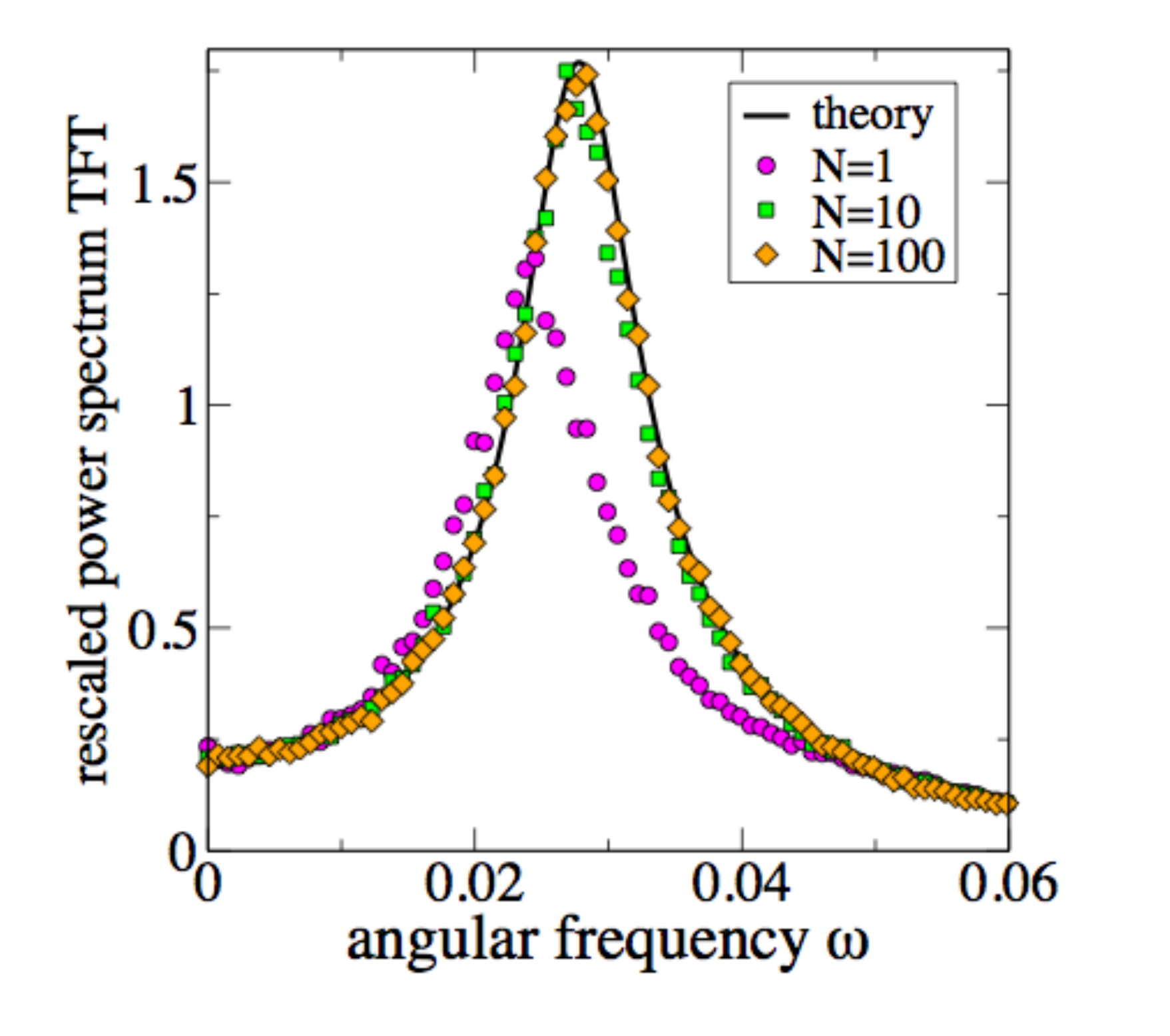}
\end{center}
% Here is how to import EPS art
\caption{\label{fig:spectrum} (Color on-line) Power spectrum of the frequency with which TFT is played. Horizontal axis shows the angular frequency $\omega$, vertical axis the spectrum of fluctuations about the deterministic fixed point. Results from numerical simulations of the stochastic dynamics are shown (markers) along with the curve predicted by the theory in the limit of large, but finite batch size. Power spectra have been re-scaled by the inverse batch size, see appendix. Model parameters are $\beta=0.1, \lambda=0.01$. Simulations are averaged over $1000$ runs.}
\end{figure}

\begin{figure*}[h]
\vspace{0em}
\begin{center}
\includegraphics[width=1.0\textwidth]{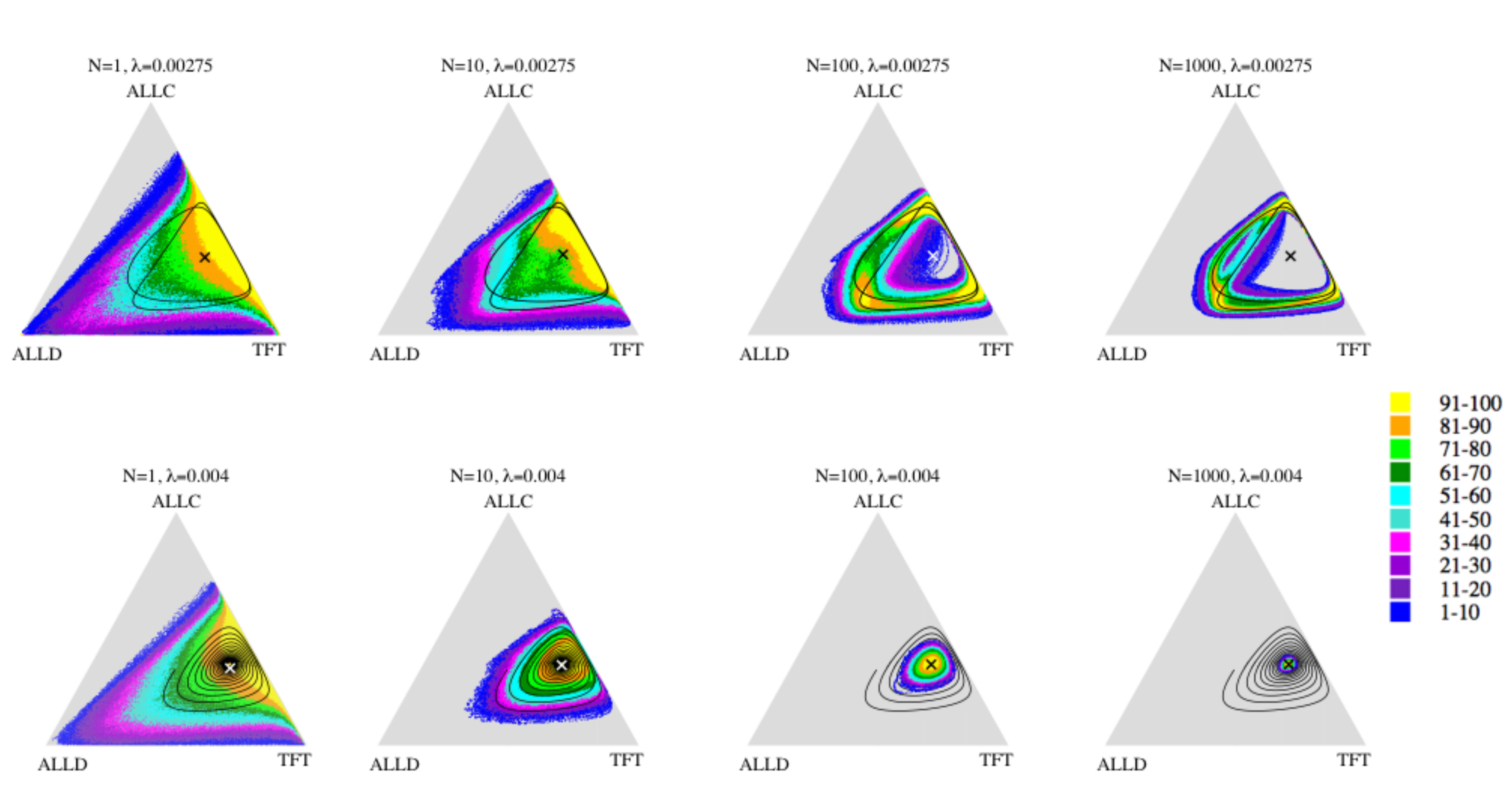}
\end{center}
% Here is how to import EPS art
\caption{\label{fig:freq} (Color on-line) Frequencies of visits of the stochastic learning dynamics.  Crosses mark the time average of the stochastic dynamics, black lines the trajectory or attractor of the deterministic discrete-time map. Data is obtained from $100$ runs of the stochastic process at an intensity of selection $\beta=0.1$. Colours indicate the frequency with which different regions are visited, a binning of strategy space is performed, and e.g. yellow stands for the $10\%$ most visited bins, orange for the next $10\%$ and so on (see legend). Grey areas are not visited by the dynamics at all in our simulations. }
\end{figure*}
 Similar noise-induced oscillation phenomena have been observed in various individual-based models of population dynamics, evolutionary game theory, epidemics and biochemical reactions, see e.g. \cite{alan,pineda,mobilia, kuske,alonso,nunes,bladon}. While the mechanism of resonant amplification in stochastic learning is analogous to the one observed in population-based models, the origin of the noise is different. In the individual-based models, large but finite populations are considered. Deterministic mean-field equations can then be derived in the limit of infinite populations. In {\em finite} populations, the dynamics remains stochastic, due to the random nature of the interactions on the microscopic level. The resulting noise scales with the inverse square root of the system size, and has been termed `demographic stochasticity' \cite{nisbet,alan}. In the learning model the source of the noise is the inaccuracy with which players sample their opponent's strategy profile at finite batch sizes, and the amplitude of the noise and of the resulting quasi-cycles is proportional to the inverse square root of the batch size $N$.

We have used a systematic expansion in the inverse batch size to characterise these cycles further (see appendix). These methods are similar to system-size expansions widely used in population-based models \cite{kampen}, even though the expansion parameter in the learning dynamics is the inverse batch size, not the size of the population. Simpler games have been studied with this technique in \cite{galla}. These expansion methods are accurate for large, but finite batch sizes. As seen in Fig. \ref{fig:spectrum} the power spectrum of the coherent oscillations can be predicted analytically with great accuracy for moderate and high batch sizes $N$. The agreement for batch sizes of $N=10$ is still reasonable,  systematic deviations are only found if the number of observations between strategy updates is reduced further.

To characterize the outcome of the stochastic learning process in more details we show the resulting stationary distributions in strategy space in Fig. \ref{fig:freq}. The panels in the upper row correspond to a memory-loss parameter for which the deterministic dynamics has a cyclic attractor. At small batch sizes stochastic learning essentially covers the entire strategy simplex, with the exception of the region near the ALLD/ALLC edge. Surprisingly, the most frequently visited points in strategy space are found along the ALLD/TFT edge, the Nash strategy ALLD is played only very rarely. At larger batch sizes the dynamics concentrates in a region about the deterministic cycle. At fast memory-loss (lower row of Fig. \ref{fig:freq}) deterministic learning has a fixed point. Again, the stochastic dynamics reaches almost the full strategy space for small batch sizes, but more and more concentration on the deterministic attractor is found as the frequency of adaptation is lowered (i.e. when the batch size is increased). In all cases shown in Fig. \ref{fig:freq} the time-average of learning is found near TFT, defection occurs only rarely.

To summarise we have here analysed in detail the learning dynamics of two fixed players interacting in a repeated prisoner's dilemma game. We find that discounting past experience in a deterministic learning produces behaviour very similar to the dynamics found in evolutionary replicator-mutator systems \cite{imhof}. Memory loss removes the stability of the ALLD fixed point, and leads to attractors near the ALLC/TFT edge of strategy space. In order to go beyond the adiabatic assumption underlying purely deterministic adaptation models, we have also addressed more realistic stochastic learning. Here, players update their strategic propensities more frequently, relying on an imperfect sampling of their opponent's strategy. We then observe persistent stochastic cycles, with a time average concentrated near TFT, paralleling earlier observations in finite evolutionary dynamics \cite{imhof}. Based on a systematic expansion technique we have characterised these cycles analytically. This method is applicable very generally, and can be used to study the effects of stochasticity in other learning models \cite{Camerer2003,fudenberg}, in machine learning problems and in algorithmic game theory \cite{Nisan2007}. Cyclic behaviour has been reported in experimental studies of multi-player learning \cite{Semmann2003}, we expect that the techniques we have introduced will be helpful in formulating and calibrating theoretical learning models describing these real-world laboratory experiments.

\begin{acknowledgments}
The author would like to thank J. D. Farmer and Y. Sato for useful discussions. This work is supported by a Research Councils UK Fellowship (RCUK reference EP/E500048/1).
\end{acknowledgments}

 \appendix
 \section*{Appendix}
\subsection{Deterministic dynamics and modified replicator equations}
\addtocounter{section}{1}

The limiting deterministic dynamics, obtained for $N\to\infty$, is given by Eqs. (3)
Taking into account Eqs. (1)  one can then write the update rule solely in terms of $\bx$ and $\by$ and finds the following map \cite{satopre}
\BE
~~~~~x_i(t+1)&=&\frac{x_i(t)^{1-\lambda}e^{\beta\sum_{j} a_{ij}y_j(t)}}{\sum_k x_k(t)^{1-\lambda}e^{\beta\sum_{j} a_{kj}y_j(t)}},\nonumber \\
~~~~~y_j(t+1)&=&\frac{y_j(t)^{1-\lambda}e^{\beta\sum_{i} a_{ji}x_i(t)}}{\sum_k y_k(t)^{1-\lambda}e^{\beta\sum_{i} a_{ki}x_i(t)}}.\label{eq:map}
\EE
Taking a continuous-time limit of (3), as discussed in \cite{satophysica}, one finds
\BE
~~~~\dot A_k&=&-\lambda A_k+\sum_{j} a_{kj}y_j,\nonumber \\
~~~~\dot B_k&=&-\lambda B_k+\sum_{i} a_{ki}x_i\label{eq:deteq}.
\EE
Using  Eqs. (1) it is then straightforward to derive deterministic continuous-time evolution equations for the frequencies $\{x_i(t), y_j(t)\}$ with which the pure strategies $i=1,\dots,S$ are played by the respective players. One finds \cite{satopre} 
\BE
\dot x_i&=&x_i\beta\left(\sum_{k}a_{ik}y_k-\sum_{k\ell} x_ka_{k\ell}y_\ell\right)-\lambda x_i\left(\log x_i-\sum_k x_k\log x_k\right),\nonumber \\
\dot y_j&=&y_j\beta\left(\sum_{k}a_{jk}x_k-\sum_{k\ell} y_ka_{k\ell}x_\ell\right)-\lambda y_j \left(\log y_j-\sum_k y_k\log y_k\right).\label{eq:sato}
\EE
These equations are occasionally referred to as the Sato-Crutchfied equations, and it is worth pointing out that they reduce to the standard replicator equations for the case of learning without memory loss ($\lambda=0$). Furthermore their behaviour is solely determined by the ratio $\lambda/\beta$. If this ratio is fixed then the role of the remaining parameter is merely to set the time scale. It is also easy to verify that the fixed points of Eqs. (\ref{eq:sato}) coincide with those of Eqs. (\ref{eq:map}). The behaviour of these dynamics can be quite intricate, depending on the structure of the underlying game. Sato et al. have for example identified chaotic motion in modified versions of the celebrated rock-paper-scissors game \cite{satopnas, satopre, satophysica}.

Fig. 1 in the main text has been obtained from a numerical integration of Eqs. (\ref{eq:sato}), using an Euler-forward scheme. We point out that it is hard to accurately determine the shape of cyclic attractors such as the one in the top-right panel of Fig. 1, even when integrating the dynamics up to large times of up to $5\cdot10^6$ and/or at small time stepping ($dt\approx 10^{-3}$). The cycle in Fig. 1 should therefore be understood as an illustration, rather than as a quantitative characterisation of the attractor.

\subsection{Analytical characterisation of stochastic cycles}\label{sec:analytical}
It is possible to make analytical progress and to compute the spectrum of the oscillations between cooperation and defection analytically in the limit of large, but finite batch sizes $N$. 

We start from the dynamics of Eq. (3),
\BE
A_k(t)&=&(1-\lambda)A_k(t)+\frac{1}{N}\sum_{\alpha=1}^N~a_{k,i_\alpha(t)}, \nonumber \\
B_k(t)&=&(1-\lambda)B_k(t)+\frac{1}{N}\sum_{\alpha=1}^N~a_{k,j_\alpha(t)},\label{eq:abbatchupdate2}
\EE
and note that the expression  $\frac{1}{N}\sum_{\alpha=1}^N~a_{k,i_\alpha(t)}$ on the right-hand side is a random variable at finite batch sizes $N$. The same is true for the analogous expression in the update rule for $B_k$. The mean value of  $\frac{1}{N}\sum_{\alpha=1}^N~a_{k,i_\alpha(t)}$  is given by $\mu_k(t)=\sum_{j} a_{kj}y_j(t)$, given that the $j_\alpha(t)$ are drawn from the mixed strategy profile $\by(t)$, i.e. action $\ell\in\{ALLC, ALLD, TFT\}$ occurs with frequency $y_\ell(t)$ on average. Similarly $\frac{1}{N}\sum_{\alpha=1}^N~a_{k,j_\alpha(t)}$ has an average of $\nu_k(t)=\sum_{i} a_{ki}x_i(t)$. Separating off fluctuations, and anticipating their scaling with $N$, we write
\BE
\frac{1}{N}\sum_{\alpha=1}^N~a_{k,i_\alpha(t)}&=&\sum_{j} a_{kj}y_j(t)+\frac{1}{\sqrt{N}}\xi_k(t),\nonumber\\
\frac{1}{N}\sum_{\alpha=1}^N~a_{k,j_\alpha(t)}&=&\sum_{i} a_{ki}x_i(t)+\frac{1}{\sqrt{N}}\eta_k(t)\label{eq:fluct}.
\EE
By means of the central limit theorem $\xi_k(t)$ and $\eta_k(t)$ can, in the limit of large but finite $N$, be approximated as Gaussian noise variables of mean zero and with the following correlations
\BE
\avg{\xi_k(t)\xi_\ell(t')}&=&\delta_{tt'}\sum_{j}\left\{y_j(t)\left[a_{kj}-\mu_k(t)\right]\left[a_{\ell j}-\mu_\ell(t)\right]\right\},\nonumber \\
\avg{\eta_k(t)\eta_\ell(t')}&=&\delta_{tt'}\sum_{i}\left\{x_i(t)\left[a_{ki}-\nu_k(t)\right]\left[a_{\ell i}-\nu_\ell(t)\right]\right\},\nonumber \\
\avg{\xi_k(t)\eta_\ell(t')}&=& 0.
\label{eq:corr}
\EE
Here $\delta_{tt'}=1$ for $t=t'$ and $\delta_{tt'}=0$ otherwise.These expressions are obtained for example by writing 
\be
\xi_k(t)=\sqrt{N}\left[\frac{1}{N}\sum_{\alpha=1}^N~a_{k,i_\alpha(t)}-\mu_k(t)\right],
\ee
followed by a straightforward evaluation of the above correlators to the appropriate order in $N^{-1/2}$, and taking into account the statistics of the $i_\alpha(t)$.

We can now proceed to insert these expressions into the map (\ref{eq:map}) and find
\BE
~~~x_i(t+1)&=&\frac{x_i(t)^{1-\lambda}e^{\beta[\sum_{j} a_{ij}y_j(t)+N^{-1/2}\xi_i(t)]}}{\sum_k x_k(t)^{1-\lambda}e^{\beta[\sum_{j} a_{kj}y_j(t)+N^{-1/2}\xi_k(t)]}},\nonumber \\
~~~y_j(t+1)&=&\frac{y_j(t)^{1-\lambda}e^{\beta[\sum_{i} a_{ji}x_i(t)+N^{-1/2}\eta_j(t)]}}{\sum_k y_k(t)^{1-\lambda}e^{\beta[\sum_{j} a_{kj}x_j(t)+N^{-1/2}\eta_k(t)]}}.\label{eq:longmap}
\EE
Given the presence of the noise terms $\xi_k(t)$ and $\eta_k(t)$, the mixed strategy profiles  $\{x_i(t), y_j(t)\}$ will be stochastic variables themselves. The next step is to self-consistently separate deterministic from stochastic contributions, and to derive a closed set of equations describing the evolution of fluctuations about the deterministic limit. To this end we write
\BE
~~~x_i(t)&=&\overline x_i(t)+\frac{1}{\sqrt{N}}\widetilde x_i(t), \nonumber \\
~~~y_j(t)&=&\overline y_j(t)+\frac{1}{\sqrt{N}}\widetilde y_j(t), \label{eq:sep}
\EE
where the quantities with overlines represent the deterministic contributions, and quantities with tildes are stochastic fluctuations. Eqs. (\ref{eq:longmap}) can be written in the form
\BE
~~~x_i(t+1)&=&f_i(\bx(t),\by(t),\boldxi(t)), \nonumber \\
~~~y_j(t+1)&=&g_j(\bx(t),\by(t),\boldeta(t)),\label{eq:comp}
\EE
with suitable functions $\{f_i,g_j\}$. One proceeds by substituting (\ref{eq:sep}) on both sides of Eq. (\ref{eq:comp}), followed by a systematic expansion in powers of $N^{-1/2}$. To lowest order one finds
 \BE
 ~~~\overline x_i(t+1)=f_i(\overline{\bx}(t),\overline{\by}(t),0), \nonumber \\
 ~~~\overline y_j(t+1)=g_j(\overline{\bx}(t),\overline{\by}(t),0),
 \EE
 i.e. one recovers the deterministic map (\ref{eq:map}).

While the calculation up to now applies to any deterministic trajectory, we will from now on restrict the discussion to an asymptotic regime, and assume that the deterministic dynamics has reached a fixed point $\overline{\bz}^*=(\overline{\bx}^*,\overline{\by}^*)$.  This is appropriate in the context of the present investigation, as we are interested in stochastic quasi-cycles about deterministic fixed points. Based on the restriction to deterministic fixed points further analytical progress is relatively straightforward \footnote{A full analytical characterisation of stochastic effects is possible also for periodic attractors of the deterministic dynamics, this has been discussed in the context of chemical reaction systems in \cite{boland} and \cite{boland2}. Such approaches are based on Floquet theory, and we expected that they are applicable also in the learning scenario (with suitable modifications to accommodate the the discrete-time dynamics). This is beyond the scope of the work presented in this paper. We point out however that all equations up to (\ref{eq:corr3}) are valid for any deterministic trajectory, provided the fixed point values $\overline{\bz}^*$ in the relevant expressions are replaced by their time-dependent counterparts.}.

In next-to-leading order of the expansion in powers of $N^{-1/2}$ one has
 \BE
 ~~~\widetilde x_i(t+1)&=&\sum_k\left(\left. \frac{\partial f_i({\bx}, {\by},\boldxi)}{\partial x_k}\right|_{({\bx}^*,{\by}^*,0)}\widetilde x_k(t)   + \left. \frac{\partial f_i({\bx}, {\by},\boldxi)}{\partial y_k}\right|_{(\overline{\bx}^*,\overline{\by}^*,0)}\widetilde y_k(t)   \right)+\kappa_i(t) \nonumber \\
  ~~~\widetilde y_j(t+1)&=&\sum_k\left(\left. \frac{\partial g_j( {\bx}, {\by},\boldxi)}{\partial x_k}\right|_{( {\bx}^*, {\by}^*,0)}\widetilde x_k(t)   + \left. \frac{\partial g_j( {\bx}, {\by},\boldxi)}{\partial y_k}\right|_{(\overline{\bx}^*,\overline{\by}^*,0)}\widetilde y_k(t)   \right)+\rho_j(t) 
 \EE
 where
 \BE
 ~~~\kappa_i(t)&=&\beta\left(\overline x_i^*\xi_i(t)-\overline x_i^*\sum_k \overline x_k^*\xi_k(t)\right),\nonumber \\
 ~~~\rho_j(t)&=&\beta\left(\overline y_j^*\eta_j(t)-\overline y_j^*\sum_k \overline y_k^*\eta_k(t)\right)\label{eq:compnoise}.
 \EE
 Writing $\bz\equiv(z_1,\dots,z_6)=(x_1,x_2,x_3,y_1,y_2,y_3)$, and using the notation $\bz(t)=\overline{\bz}^*+N^{-1/2}\bzeta(t)$ to separate deterministic from stochastic contributions ($\bzeta=(\widetilde x_1,\widetilde x_2,\widetilde x_3,\widetilde y_1,\widetilde y_2, \widetilde y_3)$) one has
\be
\bzeta(t+1)=\mathbb{J}^*\bzeta(t)+\boldphi(t),\label{eq:langevin}
\ee        
where $\mathbb{J^*}$ is the $6\times 6$ Jacobian matrix of the deterministic equations (\ref{eq:map}), evaluated at the fixed point $\overline\bz^*=(\overline\bx^*,\overline\by^*)$. The variable $\boldphi=(\varphi_1,\dots,\varphi_6)=(\kappa_1,\kappa_2,\kappa_3,\rho_1,\rho_2,\rho_3)$ represents Gaussian noise, uncorrelated in time, but with cross-correlations between the different components:
\be
\avg{\varphi_a(t)\varphi_b(t')}=\delta_{tt'}D^*_{ab}\label{eq:corr2}.
\ee
The elements of $\mathbb{D}^*$ can be expressed in terms of the deterministic variables $\overline{\bz}$. More precisely one has, using Eqs. (\ref{eq:compnoise}),
\BE
D_{ij}&=&\beta^2\bigg[\olx_i^*\olx_j^*\avg{\xi_i\xi_j}-\olx_i^*\olx_j^*\sum_{k=1}^3\olx_k^*\avg{\xi_i\xi_k}-\olx_j^*\olx_i^*\sum_{k=1}^3\olx_k^*\avg{\xi_j\xi_k}+\olx_i^*\olx_j^*\sum_{k=1}^3\sum_{\ell=1}^3\olx_k^*\olx_\ell^*\avg{\xi_k\xi_\ell}\bigg]
\label{eq:zcorr1}
\EE
and
\BE
D_{i+3,j+3}&=&\beta^2\bigg[\oly_i^*\oly_j^*\avg{\eta_i\eta_j}-\oly_i^*\oly_j^*\sum_{k=1}^3\oly_k^*\avg{\eta_i\eta_k}-\oly_j^*\oly_i^*\sum_{k=1}^3\oly_k^*\avg{\eta_j\eta_k}+\oly_i^*\oly_j^*\sum_{k=1}^3\sum_{\ell=1}^3\oly_k^*\oly_\ell^*\avg{\eta_k\eta_\ell}\bigg]
\label{eq:zcorr2}
\EE
for $i,j\in\{1,2,3\}$. The noise variables $\varphi_a$ with $a\in\{1,2,3\}$ are uncorrelated from those with $a\in\{4,5,6\}$ so that the matrix $\mathbb{D}^*$ is block diagonal ($D^*_{ab}$ and $D^*_{ba}$ both vanish if $a\in\{1,2,3\}$ and $b\in\{4,5,6\}$). The covariances of the noise variables $\{\xi_k\}$ and $\{\eta_k\}$ are given by (\ref{eq:corr}). One further potentially subtle point deserves some attention here. The covariance elements of the noise variables $\{\xi_k\}$ and $\{\eta_k\}$  as given in (\ref{eq:corr}) depend on the variables $\bz(t)=(x_1(t),x_2(t),x(t),y_1(t),y_2(t),y_3(t))$.  These in turn have deterministic and stochastic contributions, $\bz=\overline\bz^*+N^{-1/2}\bzeta$. Within our expansion in powers of $N^{-1/2}$ it is justified to self-consistently suppress the stochastic contributions $N^{-1/2}\bzeta$ to the variables $\bz$ in Eq. (\ref{eq:corr}), as these contributions would not affect results to the order of $N^{-1/2}$ we are working at. For the purposes of Eqs. (\ref{eq:zcorr1}) and (\ref{eq:zcorr2}) we therefore use
\BE
\avg{\xi_k(t)\xi_\ell(t')}&=&\delta_{tt'}\sum_{j}\left\{\oly_j^*\left[a_{kj}-\overline{\mu}_k^*\right]\left[a_{\ell j}-\overline{\mu}_\ell^*\right]\right\}\nonumber \\
\avg{\eta_k(t)\eta_\ell(t')}&=&\delta_{tt'}\sum_{i}\left\{\olx_i^*\left[a_{ki}-\overline{\nu}_k^*\right]\left[a_{\ell i}-\overline{\nu}_\ell^*\right]\right\},\nonumber\\
\label{eq:corr3}
\EE
where $\overline{\mu}^*_i=\sum_j a_{ij}\oly_j^*$ and $\overline{\nu}^*_j=\sum_i a_{ji}\oly_i^*$.

Starting from the linear equation (\ref{eq:langevin}) we now move to Fourier space and write $\widehat \zeta_a(\omega)$ for the Fourier transform of $\zeta_a(t)$ and similarly for the noise components $\varphi_a$ ($a=1,\dots,6$). One then has
\be\label{eq:inv}
\widehat \bzeta(\omega)=\mathbb{M}^{-1}\widehat \boldphi(\omega),
\ee
where $\mathbb{M}=e^{i\omega}\mathbb{I}-\mathbb{J}^*$. The notation $\mathbb{I}$ here indicates the $6\times 6$ identity matrix. The power spectra of the components of $\bzeta$ can then be obtained from Eq. (\ref{eq:inv}), taking into account  (\ref{eq:corr2}), i.e. the fact that $\avg{\widehat\varphi_a(\omega)\widehat\varphi_b(\omega')}=\delta(\omega+\omega')\mathbb{D}^*_{ab}$. One then has  
\be
P_{aa}(\omega)=\avg{|\widehat\zeta_a(\omega)|^2}=\sum_{bc}(\mathbb{M}^{-1})_{ab}\mathbb{D}^*_{bc}(\mathbb{M^\dagger}^{-1})_{ca}. \label{eq:spec}
\ee 
The right-hand-side can be evaluated numerically using the explicit form of the Jacobian $\mathbb{J}^*$ and of the noise covariance matrix $\mathbb{D^*}$. These quantities only depend on the fixed point $\overline{\bz}^*$ of the deterministic dynamics, which again can be obtained by numerical iteration of the map (\ref{eq:map}), or as a numerical solution of the corresponding fixed point relations.

Power spectra of this type are plotted in Fig. \ref{fig:spectrum}. It is important to note that these represent power spectra of the variables $\bzeta$, i.e. the pre-factor $1/\sqrt{N}$ in Eq. (\ref{eq:sep}) have already been scaled out. The theory hence predicts that these re-scaled spectra are independent of the batch size $N$, which is why the different spectra in Fig. 3 collapse on one curve (with the exception of the $N=1$ case, at these batch sizes the theory does not apply). To put it in other words: in order to obtain the raw spectra of deviations from the deterministic fixed point the amplitude of the different spectra in Fig. 3 each need to be divided by $N$. It is then clear that in absolute terms fluctuations are larger for small batch sizes (e.g. $N=1$) than for larger batches (e.g. $N=1000$). The amplitude of fluctuations and of the cycles scales as $N^{-1/2}$.

We stress at this point that amplification mechanisms have been studied extensively in population-based models, chemical reaction systems, evolutionary game theory and epidemiology, see for example \cite{alan, alonso, mobilia, bladon,nunes,pineda,kuske}. While the mechanism of amplification is similar to the one discussed here the source of the noise is different. Stochasticity in these population-based models arises when populations are finite, hence the term `demographic stochasticity' \cite{nisbet}. The approach taken in the population models is based on a systematic expansion in the inverse square root of the system size. These techniques are originally due to van Kampen \cite{kampen}. In the learning system randomness instead comes from imperfect sampling of the opponent's mixed strategy, and the expansion parameter is the inverse square root of the number of observations made between strategy updates. Similar batch-size expansions have previously been applied to simpler games in \cite{galla}.

\subsection{Comparison of continuous-time and discrete-time deterministic dynamics}
\begin{figure}[t]
\vspace{6em}
\begin{center}

\includegraphics[width=0.7\textwidth]{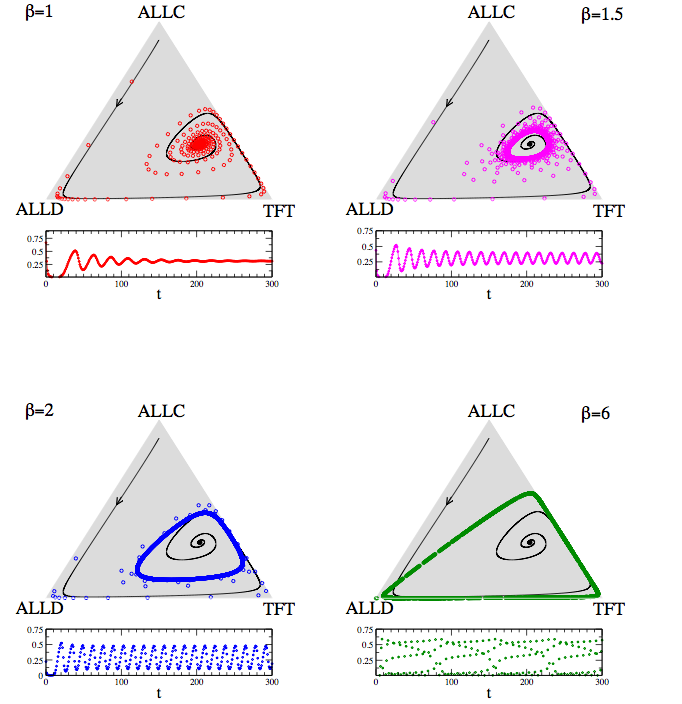}
\end{center}
% Here is how to import EPS art
\caption{\label{fig:supp1} (Color on-line) Comparison of discrete and continuous-time deterministic learning. We show trajectories of the dynamics at fixed $ \lambda/\beta=0.1$, started from homogeneous initial conditions, $\bx(t=0)=\by(t=0)$. The black line in each simplex is obtained from $\beta=0.01$, and represents the continuous-time limit. The symbols are for $\beta=1$ (upper left panel), $\beta=1.5$ (upper right), $\beta=2$ (lower left) and $\beta=6$ (lower right). In each panel we show the trajectory in the strategy simplex, as well as the corresponding time series of the propensity, $x_1(t)$ of playing ALLC.}
\end{figure}
The modified replicator equations, suggested by Sato and Crutchfield \cite{satopre} are differential equations and as such describe a continuous-time learning process. This approximation is valid for $\beta\ll 1$. The behaviour of discrete-time deterministic learning can however be quite different from this continuous-time limit, as illustrated in Fig. \ref{fig:supp1}. We here fix the ratio $\lambda/\beta$ and consider the behaviour at different values of $\beta$. For small $\beta$ the discrete-time maps behaves essentially like the continuous dynamics, and has a stable spiral fixed point. As $\beta$ is increased however, this fixed point becomes unstable, and a cyclic attractor develops\footnote{While the attractor of the dynamics appears to be a closed cyclic object, it is hard to determine numerically whether the trajectory is actually periodic, as we cannot exclude small drifts. The attractors plotted in the figure may therefore be invariant curves of the map, rather than actual cycles.}. Further increasing $\beta$ enlarges the cycle, until its the attractor finally becomes a rather large  triangular shaped object as depicted in panel d). It is here important to note that even though the attractor set looks smooth, the dynamics does not revolve around the attractor in a continuous motion. We expect that more complicated behaviour, such as chaotic attractors, will in principle be possible, even though we have not observed them for the present game and the present learning dynamics. Other learning rules in similar games have however been shown to admit chaotic motion, see \cite{ochea}.

\subsection{Inhomogeneous initial conditions}
\begin{figure}[t]
\vspace{2em}
\begin{center}
\includegraphics[width=0.9\textwidth]{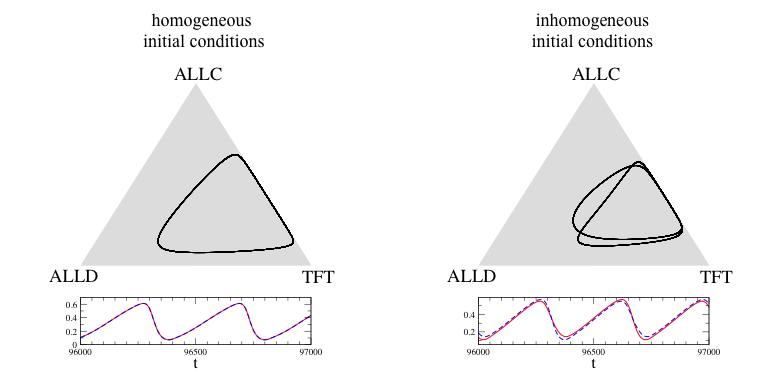}
\end{center}
% Here is how to import EPS art
\caption{\label{fig:sfig1prime} (Color on-line) Effects of initial conditions.  Left: Attractor obtained from running the deterministic map (\ref{eq:map}) starting from homogeneous initial conditions. The lower panel shows the ALLC-component of the mixed strategy of each player, the trajectories of both players are identical, $\bx(t)=\by(t)$. Right: Attractor obtained from inhomogeneous initial conditions, $\bx(t=0)\neq\by(t=0)$.  Lower panel shows that the ALLC components of both players are not identical, but that there is a relative shift in time, $\bx(t)=\by(t-\Delta t)$. Parameters are $\beta=0.01, \lambda=0.00275$ in both panels.}
\end{figure}

The map defined by Eqs. (\ref{eq:map}) describes the coupled dynamics between the two players. It is the analogue of a two-population replicator equation in evolutionary dynamics. If started from {\em homogeneous} initial conditions, $\bx(t=0)=\by(t=0)$, the deterministic map will operate in the space in which $\bx(t)=\by(t)$, i.e. both players will play identical mixed strategies. This is not generally the case for the stochastic dynamics, as the randomness in the players' decisions will break the symmetry. We find that starting the deterministic map from {\em inhomogeneous} initial conditions ($\bx(t=0)\neq \by(t=0)$) may affect the resulting attractors, see Fig. \ref{fig:sfig1prime}.

\subsection{Effect of selection intensity on stochastic dynamics}
\begin{figure}[t]
\vspace{2em}
\begin{center}
\includegraphics[width=0.8\textwidth]{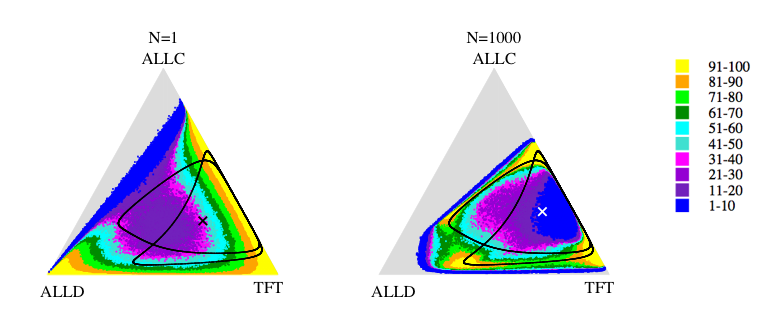}
\end{center}
% Here is how to import EPS art
\caption{\label{fig:suppfig_beta} (Color on-line) Effect of intensity of selection in stochastic dynamics. Figure shows the attractor of deterministic learning (black curves, started from inhomogeneous initial conditions) along with distributions obtained from stochastic learning at $N=1$ (left) and $N=1000$ (right) for $\beta=1$ and $\lambda=0.04$. The ratio $\lambda/\beta$ is thus as in the panels in the upper row of Fig. 4, but with the intensity of selection increased tenfold.}
\end{figure}

The role of the intensity of selection, $\beta$, on the stochastic dynamics is illustrated in Fig. \ref{fig:suppfig_beta}. The ratio $\lambda/\beta$ is the same as in Fig. 4, but with an increased value of $\beta$.  Comparing Fig. \ref{fig:suppfig_beta} with the upper panels of Fig. 4 shows that an increase of the selection intensity drives the deterministic dynamics to a cycle, and the stochastic dynamics towards the edges of the strategy simplex, especially at small batch sizes $N$ (strong noise). A behaviour not too dissimilar from that of the corresponding evolutionary system emerges, c.f. Fig. 2a and 2c of \cite{imhof}.
\subsection{Dominance of TFT in stochastic learning at low memory-loss}
\begin{figure}[t]
\vspace{2em}
\begin{center}
\includegraphics[width=0.5\textwidth]{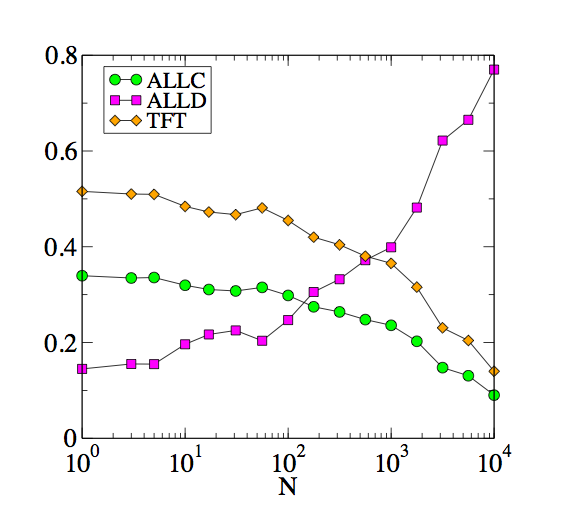}
\end{center}
% Here is how to import EPS art
\caption{\label{fig:suppfig2} (Color on-line) Temporal average of the player's mixed strategies in stochastic learning as a function of the batch size. Parameters are fixed at $\beta=0.01, \lambda=10^{-4}$. At small batch sizes (strong noise) the player's strategies are dominated by TFT, at larger values of $N$ defection is played most frequently. Data is from simulations, run for $100,000$ time steps, measurements performed in the second half of this interval, data averaged over $400$ samples.}
\end{figure}

In \cite{imhof} it was reported that evolutionary dynamics in the limit of small mutation rates chooses defection at infinite population sizes, but that a finite population of a suitable size can instead choose reciprocity (TFT). An analogous effect is seen in adaptive learning, as illustrated in Fig. \ref{fig:suppfig2}. We here choose a relatively small memory-loss rate $\lambda$, and, as a function of the batch size $N$, we measure the frequencies with which each of the three pure strategies are played asymptotically in the learning dynamics. While ALLD dominates in the deterministic limit of large $N$, TFT, i.e. reciprocity is the most frequently used pure strategy in the strongly stochastic case of small batches.
\subsection{Asynchronous updating}
\begin{figure}[t]
\vspace{2em}
\begin{center}
\includegraphics[width=0.5\textwidth]{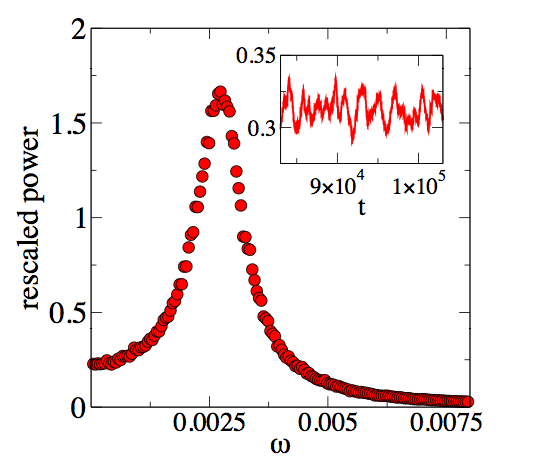}
\end{center}
% Here is how to import EPS art
\caption{\label{fig:asyn} (Color on-line) Stochastic cycles in asynchronous updating. We show the power spectrum of fluctuations of the propensity to play ALLC (main panel, averaged over $1000$ independent runs) as well as a time series from one individual run (inset). Parameters are $\beta=0.01, \lambda=0.001$, the batch size is $N=10$.}
\end{figure}
The batch dynamics assumes that both players Alice and Bob update their attractions and mixed strategies synchronously once every $N$ rounds of the game. This assumption was made mainly to simplify analytical approaches. In this section we briefly show that asynchronous updating does not alter the picture of coherent stochastic cycles. In Fig. \ref{fig:asyn} we show time series and power spectra resulting from a learning dynamics in which each player independently updates with probability $1/N$ after each individual round of the game. I.e. one round of the iterated prisoner's dilemma is played, and then for each player it is determined whether or not an update of the player's attractions and mixed strategy occurs (this happens with probability $1/N$), or whether no update is performed (this happens with probability $1-1/N$). In each update all rounds of the game since the last update are taken into account. Imposing this dynamics each player performs updates on average every $N$ iterations, but not synchronized with the other player. As seen in the figure coherent cycles are found at finite batches as before.

\subsection{Other learning rules}
In this section we briefly consider other learning rules, in particular the experienced-weighed attraction (EWA) learning proposed in \cite{Camerer2003,Ho2007}. The EWA update follows an algorithm not too dissimilar form the one discussed in the main body of this paper. In particular decisions are based on a logit rule, i.e. we have as before
\be 
x_i(t)=\frac{e^{\beta A_i(t)}}{\sum_{k}e^{\beta A_k(t)}}, ~~~~ y_i(t)=\frac{e^{\beta B_i(t)}}{\sum_{k}e^{\beta
B_k(t)}}\label{eq:camprob}.
\ee 
EWA learning uses a the following update rule for the attractions $q_i(t)$ and $r_i(t)$:
\BE
A_k(t)&=&\frac{(1-\lambda) Z(t-1) A_k(t-1)+\left[\delta+(1-\delta)I(i(t),k)\right] a_{k,j(t)}}{Z(t)} \nonumber \\
B_k(t)&=&\frac{(1-\lambda) Z(t-1) B_k(t-1)+\left[\delta+(1-\delta)I(j(t),k)\right] a_{k,i(t)}}{Z(t)}. \label{eq:camonline}
\EE
Here $i(t)$ is the action taken by player $X$ at round $t$, and $j(t)$ the action of player $y$ in round $t$. $I(\cdot,\cdot)$ indicates the Kronecker function, i.e. $I(i,j)=1$ for $i=j$, and $I(i,j)=0$ otherwise. The normalisation in the denominator is updated as

\be
Z(t)=(1-\lambda)(1-\kappa)Z(t-1)+1.
\ee
We note that $\phi$ in the notation of \cite{Camerer2003,Ho2007} is equal to $\phi=1-\lambda$ in our notation.
Eqs. (\ref{eq:camonline}) correspond to on-line learning of batch size $N=1$. One generalisation to batches of size $N$ is given by
\BE
A_k(t+1)&=&\frac{(1-\lambda) Z(t-1) A_k(t-1)+N^{-1}\sum_{\alpha=1}^N \left[\delta+(1-\delta)I(i_\alpha(t),k)\right] a_{k,j_\alpha(t)}}{Z(t)} \nonumber \\
B_k(t+1)&=&\frac{(1-\lambda) Z(t-1) B_k(t-1)+N^{-1}\sum_{\alpha=1}^N \left[\delta+(1-\delta)I(j_\alpha(t),k)\right] a_{k,i_\alpha(t)}}{Z(t)}. \label{eq:cambatch}
\EE
The parameters $\beta, \phi=1-\lambda, \kappa$ and $\delta$ have been fitted to results from real-world experiments on human subjects in \cite{Ho2007}. We here focus on the choices $\beta=1$, $\kappa=0.75$ and $\phi=0.8$, these are roughly consistent with values reported in \cite{Ho2007}, see in particular Table 4 of this reference. It is here important to note that the specific values of these parameter estimates may depend on the detailed experimental protocol, and more importantly on the game under consideration. The purpose of the present section is to show that amplified stochastic oscillations may occur in principle in EWA learning, a more detailed analysis in dependence on model parameters is left for future work. 
\subsection{The case $\delta=1$}
\begin{figure}[t]
\vspace{2em}
\begin{center}
\includegraphics[width=0.95\textwidth]{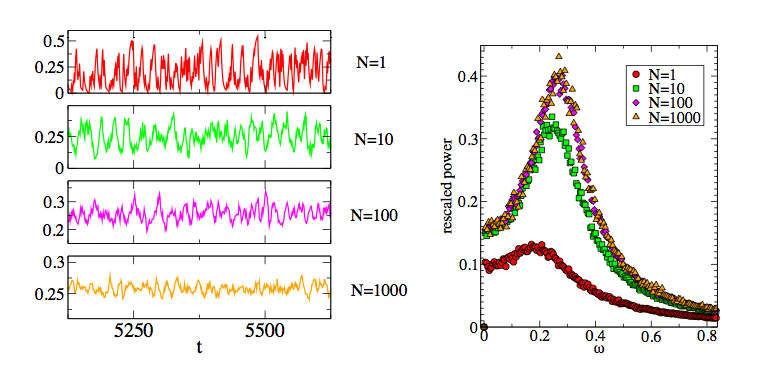} \end{center}
% Here is how to import EPS art
\caption{\label{fig:suppfigdelta1} (Color on-line) Left: Trajectories from single runs of the EWA learning process at $\beta=1,\kappa=0.75, \phi=0.8, \delta=1$. The curves show the propensity, $x_{ALLC}$ to use ALLC. Right: Corresponding re-scaled power spectra. Simulations are here averaged over at least $100$ samples.}
\end{figure}

We first investigate the case $\delta=1$, in which all foregone payoffs are re-inforced, i.e. the attractors of all pure strategies are updated, even those of strategies that have not actually been played. Results are shown in Fig. \ref{fig:suppfigdelta1}, and as seen in the figure the behaviour of the EWA learning model for these parameters is very similar to that of the simplified model of the main body of the paper. The power spectra in the right panel confirm the existence of amplified stochastic oscillations, individual trajectories at different batch sizes are shown in the left-hand panels. The spectra shown in Fig. \ref{fig:suppfigdelta1} are obtained from fluctuations about the mean of time series, and that these fluctuations have been re-scaled to take into account the $1/\sqrt{N}$ nature of their magnitude. 
\subsection{The case $\delta<1$}
\begin{figure}[t]
\vspace{2em}
\begin{center}
\includegraphics[width=0.5\textwidth]{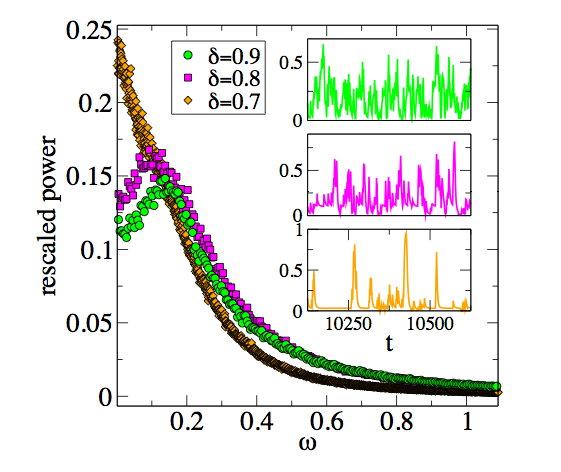}  
\end{center}
% Here is how to import EPS art
\caption{\label{fig:deltaless1} (Color on-line) The case $\delta<1$. Other parameters as in Fig. \ref{fig:suppfigdelta1}. The batch size is $N=1$. The main panel shows power spectra obtained from time series $x_{ALLC}(t)$ (averaged over $1000$ samples), the insets show trajectories $x_{ALLC}(t)$ from individual runs at $\delta=0.9,0.8,0.7$ (from top to bottom).}
\end{figure}

The case $\delta<1$ is discussed briefly in Fig. \ref{fig:deltaless1}. Here the strategic choices actually taken are re-inforced with a stronger weight than those which were not played. We find that oscillations persist, provided $\delta$ is not too small, the power spectra of fluctuations maintain their maxima at non-zero characteristic frequencies (see Fig. \ref{fig:deltaless1}). At values of $\delta$ smaller than some threshold value (which appears to depend on the other model parameters), no oscillations are found. Near $\delta=0$ the dynamics may even converge to pure actions. A further more detailed analysis of the EWA model is possible based on the techniques developed in \cite{galla} and the present paper. In particular the analytical methods of Sec. \ref{sec:analytical} will allow for a detailed study of the regions of parameter space in which cycles between co-operation and reciprocity are to be expected. This is will be the topic of future work.

\end{document}